# FCC-ee: ENERGY CALIBRATION

M. Koratzinos, A. Blondel, University of Geneva, Geneva, Switzerland; E. Gianfelice-Wendt, Fermilab[1], Batavia IL, USA; F. Zimmermann, CERN, Geneva, Switzerland.

*Abstract*

The FCC-ee aims to improve on electroweak precision measurements, with goals of 100 keV on the Z mass and width, and a fraction of MeV on the W mass. Compared to LEP, this implies a much improved knowledge of the centre-of-mass energy when operating at the Z peak and WW threshold. This can be achieved by making systematic use of resonant depolarization. A number of issues have been identified, due in particular to the long polarization times. However the smaller emittance and energy spread of FCC-ee with respect to LEP should help achieve a much improved performance.

## INTRODUCTION

Accurate energy determination is a fundamental ingredient of precise electroweak measurements. In the case of LEP1 the centre of mass energy at and around the Z peak was known with an accuracy of around 2×10$^{-5}$. The exact contribution of the energy error to the mass and the width of the Z are presented in [1].

The proposed circular collider FCC-ee [2] is capable of delivering statistics a factor ~10$^5$ larger than LEP at the Z and WW energies, therefore there is a need not only to achieve similar performance as far as energy determination is concerned, but to do significantly better.

The beam energy of large storage rings continuously changes due to internal and extraneous causes. This evolution can be modelled, but energy changes are many orders of magnitude larger than the instantaneous accuracy of a depolarization measurement. For example, small changes in the diameter of the ring due to elastic deformations of the earth's crust (due to, for instance, tidal forces) can have a big effect on the energy of the electrons and positrons. This is due to the small momentum compaction factor $\alpha_c$ which relates changes in energy to changes in the orbit length of a storage ring:

$$\frac{\Delta E}{E} = -\frac{1}{\alpha_c}\frac{\Delta L}{L} \quad (1)$$

where $L$ is the orbit length. Table 1 shows changes in energy for a $4 \cdot 10^{-8}$ circumference change (typical for tide-induced changes) for LEP and FCC-ee.

The many other effects that contribute to energy changes are discussed in [3]. None of them has a very fast changing component, so monitoring the energy every ~10 minutes would ensure a negligible extrapolation error.

The RF configuration can give rise to different energies at the IPs and for electrons and positrons, as can the slightly different orbit for the separated rings, therefore both species should be measured, something that was not done at LEP.

**Table 1:** Change in energy of a 45GeV beam for a circumference change of $4 \cdot 10^{-8}$

| Storage ring | Circumference (km) | $\alpha_c$ | $\Delta E$ (MeV) |
|---|---|---|---|
| LEP | 27 | $2 \cdot 10^{-4}$ | 9 |
| FCC-ee | 100 | $5 \cdot 10^{-6}$ | 360 |

The only method that can provide the accuracy needed is the so-called resonant depolarization technique [3], each measurement of which has an instantaneous accuracy of O(10$^{-6}$). It is based on the fact that the spin of an electron in a storage ring (in a perfectly planar machine and in the absence of solenoids) will precess $a\gamma$ times for one revolution in the storage ring, where $a$ is the anomalous magnetic moment and $\gamma$ the Lorenz factor of the electron and therefore the spin tune $\nu$ is

$$\nu = a\gamma = \frac{aE}{mc^2} = \frac{E[MeV]}{440.6486(1)[MeV]} \quad (2)$$

Deviations from the above formula are small and are discussed in [4] and [5] where they were found negligible for LEP, but should be revised in view of the much improved precision aimed at the FCC-ee.

The average of all spin vectors in a bunch is defined as the polarization vector $\vec{P}$. Therefore the average energy of a bunch can be computed by selectively depolarizing a bunch of electrons or positrons which have been polarized to an adequate level and measuring the frequency at which this depolarization occurs. Beam polarization is usually measured by laser polarimeters which exploit the spin dependence of the Compton scattering cross section. The accuracy with which the instantaneous average energy of the bunch is computed using this method is O(100KeV) – a value much smaller than the beam energy spread.

## TRANSVERSE POLARIZATION

Electron and positron beams in a storage ring naturally polarize due to the Sokolov-Ternov effect [6]. For the purposes of energy calibration, important figures of merit are the asymptotic value of polarization that can be reached and the time constant of polarization build-up.

The maximum achievable polarization value is given by the theory as

$$P_{max} = \frac{8}{5\sqrt{3}} \cong 0.924 \quad (3)$$

---

however, numerous depolarizing effects (due to for instance machine imperfections) limit this number to lower levels.

For an initially unpolarised beam the time dependence for build up to equilibrium is

$$P(t) = P_{max}[1 - \exp(-t/\tau_{pol})] \quad (4)$$

and the built up rate is (in natural units)

$$\tau_{pol}^{-1}[s^{-1}] \approx \frac{2\pi}{99} \frac{E[GeV]^5}{C[m]\rho[m]^2} \quad (5)$$

where $C$ is the circumference of the storage ring and $\rho$ its bending radius. Therefore polarization times increase with the machine circumference and decrease with energy (Table 2). The use of wigglers [7] can decrease this time as discussed further.

**Table 2:** Polarization times without the help of wigglers in the absence of imperfections

| Storage ring | Circumf. (km) | Bending radius (km) | E (GeV) | $\tau_{pol}$ (hours) |
|---|---|---|---|---|
| LEP | 27 | 3.1 | 45 | 5.8 |
| FCC-ee | 100 | 10 | 45 | 252 |
| FCC-ee | 100 | 10 | 80 | 16 |

## POLARIZATION AND ENERGY SPREAD

One important limitation on achievable polarization levels comes from the energy spread of the beam. Energy spread scales approximately like

$$\sigma_E \propto \frac{E^2}{\sqrt{\rho}} \quad (6)$$

If we extrapolate from the measurements done at LEP [8] where the maximum energy where polarization was observed was 60.6GeV (at a level of around 8%) we get the values of Table 3. Polarization at the W pair threshold (80GeV) at FCC-ee seems possible. This is in contrast of what was achieved at LEP and another input to the physics case of this unique machine. Measurements in [8] also indicated that energy spreads larger than about 52MeV lead to a significant drop of polarization levels. Detailed simulations should eventually replace the empirical approach based on the LEP experience.

**Table 3:** Extrapolation of LEP data to other machines regarding the maximum energy below which polarization levels will be adequate for depolarization measurements

| Storage ring | C(km) | Maximum energy with polarization (GeV) |
|---|---|---|
| LEP | 27 | 61 |
| FCC-ee | 80 | 80 |
| FCC-ee | 100 | 84 |

## RESONANT DEPOLARIZATION AT THE FCC-EE

The way that resonant depolarization measurements are performed is the following: Only one bunch is targeted at a time. Since the colliding rate is much larger than the polarization rate, for polarization to build up, this bunch needs to be a non-colliding bunch. It should be stated here that operation with colliding and non-colliding bunches might be a challenge due to the different tune shifts of the two species of bunches. The measurement proper consists of measuring the spin precession frequency by introducing a resonance in a 'trial and error' fashion. If no depolarization is observed, the frequency used is not the correct depolarizing frequency. The bunch remains polarized. If the bunch depolarises, the frequency corresponds to the exact mean energy of the bunch at that moment. To observe the polarization change, polarization levels of 5-10% are needed depending on the polarimeter.

## WIGGLERS

The natural polarization time for large rings is very long as seen in Table 2 (even though we only need polarization levels of 5-10%, so that we can divide the numbers in the table by a facto 10 to 20). The expected mean time between failure cannot be assumed to be more than a few hours or a day at most. A way to reduce polarization times is the use of wigglers [7]. Wigglers are dipole magnets with two parts: a low field region and a high field region so that the integral field seen by the electrons is zero. However they help, as polarization time scales with the square of the field and polarization levels are not affected provided that the wiggler asymmetry (the ratio of lengths of the positive and negative field magnets) is larger than ~5.

Wigglers have, however, two undesired effects: They increase the energy spread and they contribute to the SR power budget of the machine. Therefore a possible strategy would be to use them is such a way that the energy spread is less than some pre-determined maximum and to switch them on only where necessary.

The maximum energy spread that can be tolerated can be determined by simulation or, more pessimistically, by using the LEP experience where, as discussed earlier, was determined to be around 52MeV. In the absence of a new design, we consider the wigglers suggested for LEP [7] that have an asymmetry of 6.15 and pole lengths of 0.65m and 4m for the strong and the weak field respectively.

The polarization time and wiggler SR power dissipated for various configurations can be seen in Table 4. These results have been obtained by simulation (SLIM) and are close to the analytical calculation. In each case we have pushed the wiggler field while keeping the energy spread below 52MeV. B+ is the field of the strong pole. As can be seen, polarization times are reduced by a large factor when using wigglers. Interestingly, polarization times depend only weakly on the number of wigglers installed (but a higher field per wiggler is needed).

Therefore useful polarization levels (5-10%) are reached after 70-140 minutes. The SR power dissipated by the wigglers is rather large, although it is reduced if one operates one wiggler at a high field rather than many at a reduced field. It should be noted here that wigglers introduce more damping and might help to achieve higher beam-beam parameters, partly compensating the luminosity loss due to wiggler SR power – this is a topic that needs to be investigated.

*Wiggler operation*

A possible strategy therefore emerges: Wigglers need to be used. For the case of FCC-ee, 250 non-colliding bunches are sufficient. The wigglers can be switched on as soon as the machine starts filling up and can be switched off when 5-10% polarization is achieved. Machine fill-up times are expected to be around 30 minutes, therefore an extra ~50-100 minute dead time is introduced while polarization builds up and during which period no meaningful energy measurement can be performed. Also, due to the power taken up by the wigglers, the luminosity of the machine will be lower than during normal operation. Physics studies which do not need precise energy determination can take place, though.

When the required level of polarization for the non-colliding bunches has been achieved, the wigglers can be turned off and the depolarization measurements can start. Measuring and replacing 5 bunches for 5 depolarization measurements per hour, the FCC-ee will exhaust all non-colliding bunches in 50 hours, during which time the used non-colliding bunches will have been polarized again to more than 10%. We will investigate if wiggler operation at a reduced setting during physics could be beneficial to the energy determination or overall performance. Also, the study of collimating the large amount of radiation from the wigglers will be a priority.

We here assume that the number of electrons in a non-colliding bunch would be similar to the number of electrons of a normal (colliding) bunch. For the FCC-ee this number is $\sim 1.8 \cdot 10^{11}$ (similar to the LEP1 value). Having 250 out of 16700 bunches not colliding leads to an inefficiency of 1.5%.

**Table 4:** The effect of the use of wigglers on polarization times, energy spread and wiggler power dissipation according to the SLIM simulation and for the wiggler design described in **[7]**. B+ is the magnetic field of the short (strong) dipole of the wiggler.

| Machine | Energy (GeV) | No. of wigglers | B+ (T) | $\tau_{pol}$ (hours) | $P_\infty (\%)$ | $\tau_{10\%}$ (hours) | Energy spread (MeV) | Wiggler SR power/beam (MW) |
|---|---|---|---|---|---|---|---|---|
| TLEP | 45 | 0 | 0 | 252 | 92.4 | 27.3 | 17 | 0 |
| TLEP | 45 | 12 | 0.62 | 24.1 | 88.1 | 2.7 | 50 | 15 |
| TLEP | 45 | 1 | 1.35 | 27.6 | 88.1 | 3.1 | 50 | 7 |

## SIMULATION

Polarization is a strong function of machine misalignment and non-linear calculations are mandatory for evaluating the effect of the energy spread in presence of machine imperfections. Two codes are currently used. SLIM [9] is used for fast linear calculations and SISTROS [10], which has second order orbit description and non-linear spin motion, for accurate results. The 100 km ring is made out of $60^0$ FODO cells with non-dispersive insertions for wigglers. The effect of one wiggler with B+=1.35 T and of random vertical misalignment of quadrupoles ($y\delta^y_{RMS} = 200\mu m$) has been considered. The orbital tunes are Q$_x$=181.124, Q$_y$=183.207 and Q$_s$=0.117. A beam position monitor and a vertical corrector is located next to each vertical focusing quadrupole. The vertical orbit has been corrected by using either 110 correctors (MICADO algorithm) or all available correctors (SVD). In addition, in the first case the polarization axis distortion has been corrected by tuning 8 harmonic bumps [11]. Figure 1 shows polarization versus spin tune for different configurations. The increased energy spread has a large impact on polarization in presence of machine imperfections. More simulations by using the actual optics are needed for assessing in addition the impact of other error sources and of BPMs errors. However it is clear that well planned state-of-the-art correction schemes will be needed.

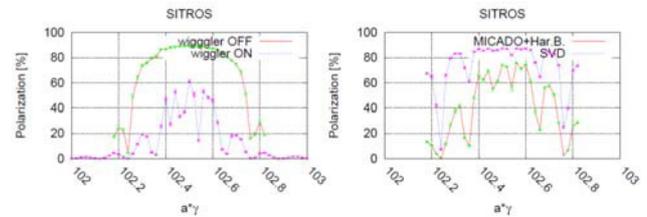

Figure 1: Polarization in presence of misalignments Left: w/o and with wiggler after correcting the closed orbit with 110 correctors. Right: wiggler on and in addition correcting the polarization axis distortion, or, after correcting the closed orbit alone with all correctors (SVD).

## CONCLUSIONS

For FCC-ee, the resonant depolarization method seems accessible at the Z (45GeV) and W (80GeV) energies. Non-colliding bunches are mandatory for the measurement. Both lepton species should be measured. Long polarization times necessitate the use of wigglers, which however are needed only during a short period at the beginning of a fill. Measurements should be performed routinely at a rate of a few per hour.